%% file: main.tex
\documentclass[aps,twocolumn,amssymb,amsmath,floats,showpacs,pre]{revtex4-1}
\usepackage{bm}
\usepackage{graphicx}
\usepackage{xcolor}
\usepackage[toc,page]{appendix}
\usepackage{comment}
\usepackage[unicode=true]{hyperref}
\usepackage{enumerate}
\usepackage{dsfont}
\usepackage{amsmath}
\usepackage{soul}
\usepackage{orcidlink}
\usepackage{enumitem}

\usepackage[normalem]{ulem}
\usepackage[ruled,vlined]{algorithm2e}
\SetKwProg{Fn}{Function}{}{}
\SetKwInOut{Input}{Input}
\SetKwInOut{Output}{Output}
\SetKwInOut{Parameter}{Parameter}
\usepackage{multirow} 
\usepackage{lineno} 
\graphicspath{{images/}}
\begin{document}

\title{Joint Trust–Distrust Structure Shapes Belief Propagation and Collective Dynamics}

\author{Mozhgan Khanjanianpak\orcidlink{0000-0003-1625-004X}}
%\email{Corresponding author: khanjanianpak@gmail.com}

\author{Yu-Ru Lin$^{1,}$}
\email{yurulin@pitt.edu}

%\vspace{0.9cm} % Adds vertical space between authors and affiliations

\affiliation{$^{1}$ \mbox{School of Computing and Information, University of Pittsburgh, Pittsburgh, PA, USA}}
%\affiliation{$^{*}$ \mbox{yurulin@ pitt.edu}}

%%%%%%%%%%%%%%%%%%%%%%%%%%%%%%%%%%%%%%%%%%%%%%
\begin{abstract}
	The propagation of beliefs in social networks depends on how individuals evaluate their relationships with others, particularly whether information from a neighbor is sufficiently trustworthy to influence their own beliefs. Distrust toward the same neighbor can, however, constrain such influence, making the joint organization of trust and distrust relevant to belief propagation. Many existing approaches represent trust and distrust as opposite or complementary aspects of a single relational quantity, thereby overlooking how their joint structure determines which relationships can effectively transmit beliefs. Here, we develop a network model in which trust and distrust are treated as distinct but statistically coupled quantities. Their joint values define the net trust and uncertainty degree of each relationship, and a neighbor is influential only when both corresponding threshold conditions are satisfied. We show that the correlation between trust and distrust substantially reshapes the consensus--fragmentation transition by controlling the availability of influential neighbors. Network connectivity further affects this transition, whereas degree heterogeneity has a more limited effect for networks with comparable mean degree. We further show that the availability of influential neighbors does not by itself determine the final consensus: when relational properties are correlated with network degree, the influence conditions can preferentially select agents with different degree compositions, thereby affecting the direction and magnitude of the collective belief. Finally, we introduce a comfort index that quantifies the expected availability of influential neighbors and analytically connects the joint trust--distrust structure to collective dynamics. Overall, the model provides a mechanistic link from the joint organization of trust and distrust, through the availability and composition of influential relationships, to the emergence of collective belief states.
\end{abstract}

\maketitle

\input{010_introduction}

\input{020_relatedwork}
\input{030_method}

\input{040_results}

\input{050_analysis}
\input{060_discussion}

\bibliography{TrustBibFile.bib}

\input{070_appendix}

\end{document}

%% file: 010_introduction.tex
\section{\label{secIntro}Introduction}

Opinion formation is a fundamental process underlying social dynamics and collective behavior. A broad range of models has been developed to describe how individuals update their beliefs through social interactions \cite{noorazar2020recent, galesic2021integrating}, including the DeGroot model of weighted averaging \cite{degroot1974reaching}, the Sznajd model of collective influence \cite{sznajd2000opinion}, the voter model of opinion spreading \cite{holley1975ergodic}, and models incorporating discrepancies between private beliefs and public expressions \cite{kuran1991now}. Among these approaches, bounded-confidence (BC) models have provided an important framework for studying how selective social influence can lead to consensus, clustering, or fragmentation \cite{deffuant2000mixing, rainer2002opinion, bernardo2024bounded}. In BC models, individuals do not necessarily interact with or accept the opinions of all their neighbors; instead, social influence is restricted according to prescribed conditions on interpersonal or opinion differences \cite{proskurnikov2017tutorial, proskurnikov2018tutorial}. This framework has subsequently been extended to account for signed interactions, algorithmic bias, and heterogeneous activity levels \cite{lorenz2007continuous, altafini2018signed, sirbu2019algorithmic, li2023bounded}.

While selective influence is central to opinion dynamics, the ability of one individual to influence another also depends on the nature of their social relationship. Trust and distrust are therefore important factors in determining whether information from a particular neighbor is considered influential \cite{castaldo2010meaning, sapp2019trust}. Previous studies have incorporated trust into opinion dynamics through different mechanisms \cite{posten2021trust, xiao2021dangers, lazer2018science}, including evolutionary trust \cite{zhang2022consensus, shen2024opinion}, similarity-based trust \cite{du2021consensus, liu2024dynamic}, and trust associated with opposing opinions \cite{jiang2024analysis}. Distrust has likewise been associated with the rejection of information and with the emergence of fragmentation and polarization \cite{ishii2021significant, primiero2017trust}. These studies demonstrate that interpersonal relationships can substantially modify the propagation and evolution of opinions. However, a common challenge is how to represent trust and distrust when they are not mutually exclusive aspects of a relationship.

Several approaches describe trust and distrust using positive and negative edge weights or signed networks \cite{kou2021building, fang2022integrating, zha2023managing}. In such representations, the sign of a relationship provides the primary distinction between positive and negative social ties. Structural-balance approaches provide another formulation in which agents are divided into groups connected predominantly by trust within groups and distrust between groups \cite{cartwright1956structural, shi2020leader, zha2023managing}. Other formulations introduce trust and distrust as separate quantities \cite{six2023distrust, aghdam2020uncertainty}. These approaches have provided valuable insight into the effects of positive and negative social relationships, but they do not necessarily capture situations in which trust and distrust coexist within the same interpersonal relationship. In real social interactions, however, these two dimensions need not be mutually exclusive: an individual may simultaneously trust another person in some respects while also distrusting them in others \cite{zhang2024profiling,  alejandro2024trust}. Thus, a representation that distinguishes the relative balance between trust and distrust from their simultaneous presence may provide a more flexible description of interpersonal relationships.

Kawabe et al. introduced such a two-dimensional representation using a fuzzy-set concurrent rating method, in which trust and distrust are represented simultaneously \cite{kawabe2019trust}. This framework provides two complementary quantities: the net balance between trust and distrust and a measure of their simultaneous presence, referred to here as the uncertainty degree. Importantly, the representation distinguishes a lack of sufficient trust from complete distrust, while also allowing trust and distrust to coexist. This makes it suitable for investigating how the joint state of a relationship affects the possibility of social influence. In particular, rather than assigning influence solely according to a single positive or negative relationship value, one can ask whether a neighbor has sufficient net trust while maintaining an acceptable degree of trust--distrust contradiction.

Motivated by this perspective, we investigate how the simultaneous presence and statistical relationship of trust and distrust affect belief propagation and the transition between consensus and fragmentation. To this end, we consider a society of agents connected through a directed acquaintance network, where each directed relationship is characterized by both a trust degree and a distrust degree. We use the two-dimensional representation of Kawabe et al. \cite{kawabe2019trust} to obtain the net trust and uncertainty degree of each relationship. A neighbor can influence an agent only when these two quantities simultaneously satisfy prescribed trust and contradiction conditions. We refer to such a neighbor as an \textit{influential neighbor}. This formulation allows us to examine belief propagation in terms of the availability and organization of relationships that are capable of transmitting influence.

We first investigate how the correlation between trust and distrust affects the consensus--fragmentation transition. By introducing a correlation parameter into the joint generation of trust and distrust, we examine how their statistical dependence modifies the accessible region of the net-trust--uncertainty plane and, consequently, the range of threshold conditions that permit belief propagation. We then examine whether these effects depend on network structure by considering homogeneous and heterogeneous networks, including Erdős--Rényi and Barabási--Albert networks. We further investigate degree-dependent relationships and initial beliefs to distinguish the effects of network connectivity from the effects of which agents are preferentially positioned to influence others. In particular, we consider cases in which trust and distrust values are preferentially associated with agents of different degrees, as well as cases in which highly connected agents are assigned biased initial beliefs.

Finally, we develop a geometric and analytical characterization of influential-neighbor availability. In the noise-free limit, the admissible trust--distrust relationships form a deterministic set in the net-trust--uncertainty plane, providing a geometric necessary condition for the availability of influential neighbors. For finite noise, we derive the joint distribution of the net trust and uncertainty degree and use it to quantify the fraction of relationships satisfying the influence conditions. This quantity, together with network connectivity, leads to a comfort index that characterizes the expected availability of influential neighbors. The resulting analysis provides a connection between the statistical structure of trust and distrust at the level of individual relationships and the collective emergence of consensus or fragmentation.

The remainder of the paper is organized as follows. Section~\ref{secBackground} reviews the background and related work on opinion dynamics, trust and distrust, and existing approaches relevant to the representation of simultaneous trust and distrust. Section~\ref{secModel} then introduces the proposed model for belief propagation, including the two-dimensional representation of trust and distrust, the influence conditions, and the construction of correlated trust--distrust relationships. Section~\ref{secResults} presents the numerical results for different correlation structures, network topologies, and degree-dependent biases. Section~\ref{secAnalysis} develops the geometric and analytical characterization of the consensus--fragmentation transition and the comfort index. The main findings and their implications are discussed in the final section.

%% file: 020_relatedwork.tex
\section{\label{secBackground}Background and related work}

\subsection{Trust, distrust, and untrust in networked systems}

Trust is commonly treated as a positive relation through which an agent accepts information or relies on another agent, whereas distrust is often regarded as its negative counterpart. This view has traditionally led to representing trust and distrust as opposite ends of a single continuum \cite{rotter1971generalized}. Within such a representation, cooperation is associated with trust, while a lack of cooperation is often interpreted as distrust. However, this binary interpretation does not distinguish between actively distrusting another agent and simply having insufficient trust. The latter situation has been referred to as \emph{untrust}, in which the available level of trust is not sufficient to support cooperation or reliance \cite{lewicki1998trust,marsh2005trust}.

This distinction is particularly relevant in networked systems, where agents rely on information obtained from their neighbors or other connected agents. Trust can facilitate information propagation, whereas distrust may prevent or interfere with such propagation. At the same time, an agent may neither sufficiently trust nor actively distrust another agent. Such intermediate states can affect how information is transmitted through a network and motivate a representation in which trust and distrust are not necessarily mutually exclusive. Accordingly, trust propagation, interference, and distrust blocking have been studied in the context of uncertain and autonomous systems \cite{marsh2005trust,ziegler2005propagation,primiero2017trust}. In this work, we build on a two-dimensional representation in which trust and distrust are treated as coupled but distinct quantities, following the framework introduced by Marsh and Dibben and further developed by Kawabe et al. \cite{marsh2005trust,kawabe2019trust}.

%%%%%%%%%%%%%%%%%%%%%
\begin{figure}[t!]
	\includegraphics[scale=0.5]{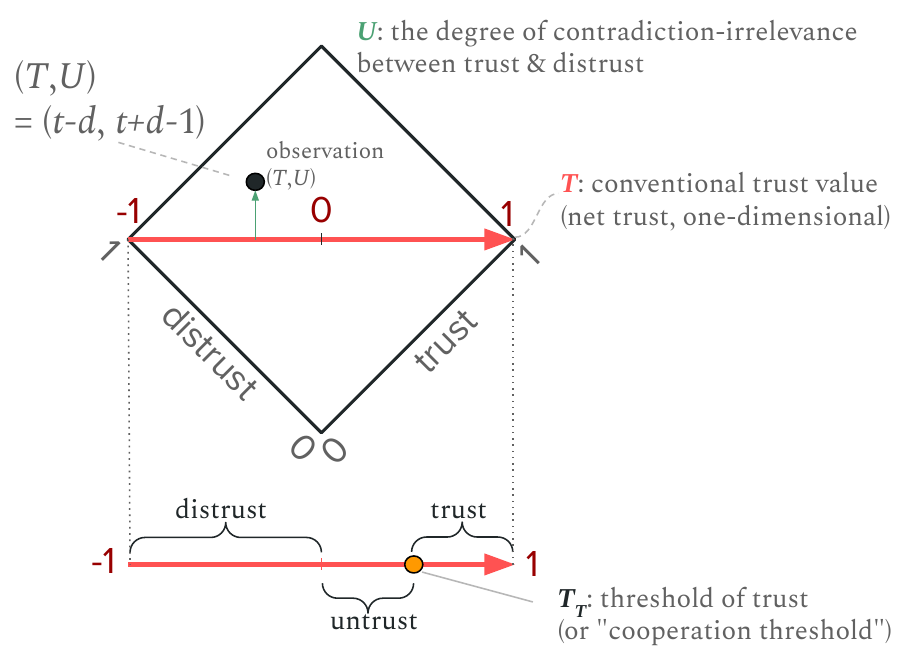}
	\centering
	\caption{\textbf{Representation of trust, distrust, and untrust}. }
	\label{fig:trust2d}
\end{figure}
%%%%%%%%%%%%%%%%%%%%%%%%%%%
%====================================
\subsection{Representation of co-existing trust and distrust}

Marsh and Dibben \cite{marsh2005trust} distinguish four related notions of trust: \emph{trust}, \emph{distrust}, \emph{untrust}, and \emph{mistrust}. Trust indicates that the trustee is regarded as trustworthy, whereas distrust indicates that the agent expects the trustee to act against its interests. In contrast, untrust describes a situation in which the level of trust is positive but insufficient to support cooperation. Mistrust refers to a change in the trust relationship following, for example, a betrayal of previously established trust or the overcoming of an initial distrust. This distinction emphasizes that the absence of trust does not necessarily imply distrust.

Following this framework, let $\tau$ and $\delta$ denote the degrees of trust and distrust, respectively, with both quantities taking values between zero and one. The conventional one-dimensional trust value can then be expressed as the net trust $T=\tau-\delta$. As illustrated in Fig.~\ref{fig:trust2d}, the line connecting $(1,0)$ and $(0,1)$ corresponds to the conventional representation in which trust and distrust are implicitly treated as complementary quantities, satisfying $\tau+\delta=1$. In this representation, a sufficiently positive net trust indicates trust, whereas a negative net trust indicates distrust. More specifically, for a given cooperation threshold $T_T$, $T\geq T_T$ corresponds to a sufficiently trusted relationship, $T<0$ to distrust, and the intermediate range $0\leq T<T_T$ to untrust. Thus, untrust represents a positive but insufficient level of net trust rather than an active state of distrust.

The two-dimensional representation additionally captures the extent to which trust and distrust coexist. This is quantified by the contradiction measure $U=\tau+\delta-1$. Positive values of $U$ indicate that both trust and distrust are relatively high, corresponding to a state of confusion, whereas negative values indicate that both are relatively low, corresponding to irrelevance. Importantly, the same value of $T$ can be obtained from different combinations of $\tau$ and $\delta$ and therefore does not uniquely determine the degree of contradiction between them. The additional variable $U$ consequently provides information that is absent from the conventional one-dimensional representation.

Kawabe et al. \cite{kawabe2019trust} further distinguish two forms of untrust within this two-dimensional representation. \emph{Confusional trust} (untrust-confusional) corresponds to $0<T<T_T$ and $U\geq0$, where trust and distrust coexist at relatively high levels. In contrast, \emph{ignorant trust} (untrust-ignorant) corresponds to $0<T<T_T$ and $U<0$, where both trust and distrust are relatively low. Building on this distinction, we introduce a contradiction threshold $U_T$ in the subsequent model. The threshold specifies the maximum level of trust--distrust contradiction that can be tolerated for an agent to be influenced. Thus, the two-dimensional representation provides the conceptual basis for incorporating not only the net balance between trust and distrust, but also their simultaneous presence, into the dynamics of belief propagation.

%% file: 030_method.tex
\section{\label{secModel} Method and model for belief propagation }

We consider a society of $N$ agents whose acquaintance network remains fixed over time and is represented by a directed adjacency matrix $A$. A directed edge from agent $j$ to agent $i$ is associated with a pair of constant values $w_{ji}=(\tau_{ji},\delta_{ji})$, where $\tau_{ji},\delta_{ji}\in[0,1]$ denote the degrees of trust and distrust that agent $i$ has in agent $j$, respectively. Following the two-dimensional representation introduced by Kawabe et al.~\cite{kawabe2019trust}, we characterize this relationship by the net trust and the degree of trust--distrust contradiction, hereafter referred to as the uncertainty degree, as follows
\begin{eqnarray}
	T_{ji} = \tau_{ji} - \delta_{ji}, 
	\label{TEqu} \\
	U_{ji} = \tau_{ji} + \delta_{ji} -1,
	\label{UEqu}
\end{eqnarray}
respectively, where both $T_{ji}$ and $U_{ji}$ take values in the interval $[-1,1]$.
 
Each agent $i$ is assigned a continuous belief variable $b_i(t)$ representing the degree to which agent $i$ considers a conceptual entity $X$ to be true at time $t$. At $t=0$, the initial beliefs are drawn from a uniform distribution over the interval $[-5,5]$. Subsequently, an agent can be influenced by its neighbors only when the corresponding trust relationship satisfies two conditions as  
\begin{enumerate}[label=(\Roman*)]
	\item the net trust of agent $i$ in agent $j$ must exceed a prescribed trust threshold, $T_{ji}>T_T$,
	
	\item the uncertainty degree must remain below a prescribed uncertainty threshold, $U_{ji}<U_T$.
\end{enumerate}
We refer to these two conditions as the influence conditions. When both conditions are satisfied, agent $j$ is an \textit{influential neighbor} (IN) of agent $i$. The beliefs of all agents are then updated synchronously according to
\begin{equation}
	\frac{db_i(t)}{dt} = \sum_{j:\,(I)\land(II)} A_{ji}[ b_j(t) - b_i(t) ], \\
	\label{andEqu}
\end{equation}
where $(I)$ and $(II)$ denote the two influence conditions above. Thus, the contribution of agent $j$ to the belief dynamics of agent $i$ is activated only when $i$ has a sufficiently high level of net trust in $j$ and the associated uncertainty degree is sufficiently low. The thresholds $T_T$ and $U_T$ therefore determine which network connections can transmit influence.

\subsection{Generative network of correlated trust and distrust}
\label{secModel_01}

We first consider trust and distrust values as independent random variables. We then introduce correlations between them to examine how their statistical relationship affects the resulting belief dynamics. In realistic settings, trust and distrust may be negatively correlated: a higher level of trust in an agent can be accompanied by a lower level of distrust toward the same agent. We characterize this relationship by a correlation parameter $\rho\in[-1,0]$ and generate $\delta_{ji}$ from a given $\tau_{ji}$ according to
\begin{equation}
	\delta_{ji}
	=	\rho(\tau_{ji}-0.5)+0.5+\epsilon,
	\label{corrEqu}
\end{equation}
where $\epsilon$ is a Gaussian noise term with zero mean and standard deviation $\sigma_\epsilon$.

For a given $\rho$ and $\sigma_\epsilon$, the trust and distrust values are generated as follows. First, $\tau_{ji}$ is drawn uniformly from $[0,1]$. A noise value $\epsilon$ is then drawn from a normal distribution with mean zero and standard deviation $\sigma_\epsilon$, and $\delta_{ji}$ is calculated from Eq.~(\ref{corrEqu}). If the resulting $\delta_{ji}$ falls outside the interval $[0,1]$, a new noise value is drawn and the procedure is repeated until a valid value is obtained. This rejection procedure ensures that both trust and distrust degrees remain within their prescribed range.

For a given adjacency matrix, the resulting values of $\tau_{ji}$ and $\delta_{ji}$ are assigned to each directed connection. We then fix the trust and uncertainty thresholds, $T_T$ and $U_T$, and numerically integrate the belief dynamics in Eq.~(\ref{andEqu}) using the Euler method with a time step of $dt=0.001$. The dynamics are integrated up to a prescribed maximum simulation time $t_{\max}$, chosen sufficiently large for the system to reach a stationary configuration. We verified that extending the simulation beyond $t_{\max}$ does not produce appreciable changes in the belief values. The final configuration at $t=t_{\max}$ is then used to characterize the resulting pattern of belief formation, corresponding either to consensus or to a persistent fragmented state.

\subsection{Generative network with heterogeneous structures}
\label{secModel_02}

We investigate belief propagation in three network structures: complete networks, Erdős--Rényi (ER) random networks, and heterogeneous networks generated using the Barabási--Albert (BA) model. The latter allows us to examine how degree heterogeneity and the presence of highly connected agents (hubs) affect the propagation of beliefs.

To investigate the role of the relationship between network connectivity and trust, we first consider a degree-dependent assignment of trust and distrust. We initially distribute the correlated pairs $(\tau_{ji},\delta_{ji})$ among the neighboring agents as described above. We then introduce a swapping probability $p_s$ to progressively align trust with the degree of the trustee. For each agent $i$, pairs of its neighbors $j_1$ and $j_2$ are considered. If a random number is smaller than $p_s$ and $(k_{j_1}-k_{j_2})(\tau_{j_1i}-\tau_{j_2i})<0$, the corresponding trust and distrust values are swapped, where $k_j$ denotes the degree of agent $j$. This procedure preserves the set of trust and distrust values assigned to the neighbors of each agent while progressively increasing their association with network degree.

When $p_s=0$, no swapping occurs and the original random assignment is retained. As $p_s$ increases, higher-degree agents are progressively more likely to receive larger trust values and, because $\tau$ and $\delta$ are correlated, lower distrust values. Thus, $p_s$ provides a control parameter for the degree--trust association while preserving the initially generated values.

Finally, we examine the effect of degree-dependent initial beliefs. In particular, we consider two types of hubs: radical hubs and moderate hubs. Starting from uniformly distributed initial beliefs, we identify the $h$ highest-degree agents as hubs and exchange their initial beliefs with those of $h$ agents selected according to the desired bias. For radical hubs, the selected beliefs are those closest to an extreme value, $+5$ or $-5$, whereas for moderate hubs they are those closest to $b=0$. In this way, the overall initial belief distribution is preserved while its assignment becomes correlated with network degree. This procedure allows us to isolate the effect of concentrating particular initial belief values on highly connected agents.

%%%%%%%%%%%%%%%%%%%%%
\begin{figure*}[t!]
	\includegraphics[scale=0.5]{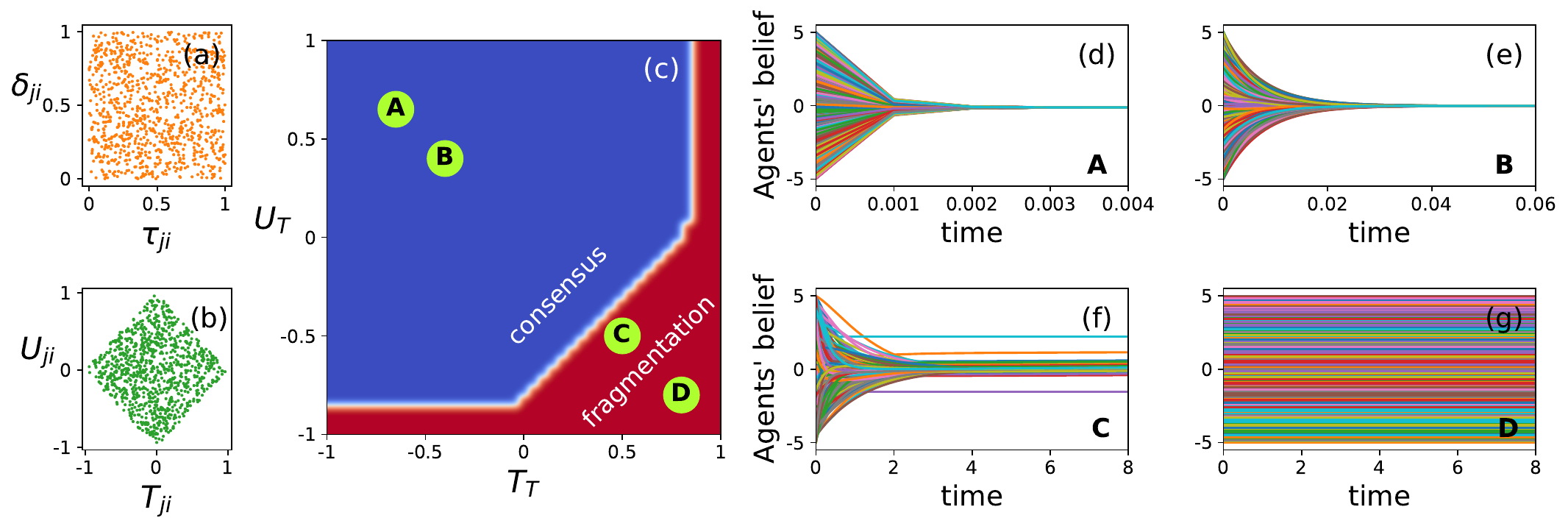}
	\centering
	\caption{\textbf{Results for uncorrelated trust and distrust values on a complete graph.} (a) Uniformly distributed trust-degree $\tau_{ji}$ and distrust-degree $\delta_{ji}$, and (b) the corresponding net trust $T_{ji}$ and uncertainty $U_{ji}$ values obtained from Eqs.~(\ref{TEqu})--(\ref{UEqu}), for agent $i$ toward its neighbors $j$ in a complete graph of size $N=1000$. (c) Phase diagram in the $(T_T,U_T)$ parameter space, showing the consensus and fragmentation regimes in blue and red, respectively. The white solid line marks the boundary separating the two phases. (d)--(g) Time evolution of agents' beliefs for the four representative parameter sets A--D indicated in panel (c), with A and B corresponding to consensus and C and D to fragmentation. Each curve represents the belief $b_i$ of one agent. Note that the time axes have different scales.}
	\label{FullyRes}
\end{figure*}
%%%%%%%%%%%%%%%%%%%%%%%%%%%

\subsection{Comfort index}
\label{secModel_03}

The two influence conditions defined for Eq.~(\ref{andEqu}) determine whether a neighbor $j$ can affect the belief of agent $i$. For a given pair of thresholds $(T_T,U_T)$, the fraction of neighboring agents satisfying these conditions depends on the joint distribution of the trust degree and distrust degree, and therefore on the correlation parameter $\rho$ and the noise level $\sigma_{\epsilon}$. We use this fraction to quantify the availability of influential neighbors (INs). Let
\begin{equation}
	F=F(T_T,U_T,\rho,\sigma_{\epsilon})
	\label{Fequ}
\end{equation}
denote the probability that a randomly selected outgoing neighbor satisfies both influence conditions, $T_{ji}>T_T$ and $U_{ji}<U_T$. For an agent $i$ with out-degree $k_i^{out}$, the corresponding expected number of INs is defined as the \textit{comfort index},
\begin{equation}
	\kappa_i=k_i^{out}F(T_T,U_T,\rho,\sigma_{\epsilon}).
	\label{kappaEqu}
\end{equation}
Thus, the comfort index quantifies the expected availability of INs for a given agent. Its dependence on network degree is explicit, whereas the function $F$ is determined by the trust--distrust statistics and the two thresholds. The analytical expression for $F$ and its derivation are given in Appendix~\ref{AppB}. To characterize the availability of INs at the population level, we further define the aggregate comfort index as
\begin{equation}
	\kappa_A=\sum_{k=k_{\min}}^{k_{\max}}NP_k\kappa_k,
	\label{kappaAEqu}
\end{equation}
where $P_k$ denotes the degree distribution. Using Eq.~(\ref{kappaEqu}), this quantity can be written as
$\kappa_A=NF\langle k^{out}\rangle$, where $\langle k^{out}\rangle$ is the mean out-degree of the network.

%% file: 040_results.tex
\section{\label{secResults}Results} 
%=======================================
\subsection{Complete graph} \label{secResults_01}

\subsubsection{Uncorrelated trust and distrust}

We first consider a complete graph of size $N$ as the simplest network structure, which also provides a convenient setting for comparison with the analytical results. Figure~\ref{FullyRes}(a) shows uniformly distributed and uncorrelated values of the trust degree $\tau_{ji}$ and distrust degree $\delta_{ji}$ assigned by agent $i$ to its neighbors $j$. Applying Eqs.~(\ref{TEqu}) and~(\ref{UEqu}) yields the corresponding net trust $T_{ji}$ and uncertainty degree $U_{ji}$, shown in Fig.~\ref{FullyRes}(b). We then examine how the trust threshold $T_T$ and uncertainty threshold $U_T$ determine the transition between consensus and fragmentation.

Figure~\ref{FullyRes}(c) shows the phase diagram in the $(T_T,U_T)$ parameter space. For $T_T<0$, the system reaches consensus over most of the range of $U_T$, except for very low uncertainty thresholds ($U_T\lesssim -0.87$). For $T_T>0$, the critical uncertainty threshold $U_T^c$ separating consensus from fragmentation increases approximately linearly with $T_T$. At sufficiently large trust thresholds ($T_T\gtrsim0.87$), consensus is no longer reached even at the maximum uncertainty threshold, $U_T=1$. Thus, sufficiently restrictive trust requirements prevent the system from reaching global consensus regardless of the allowed level of trust--distrust uncertainty. Notably, the consensus phase occupies a substantially larger portion of the parameter space than the fragmentation phase. This feature is related to the assumption of uncorrelated trust and distrust degrees used here. Since trust and distrust degrees are generally expected to be negatively correlated, introducing such a correlation can substantially modify the accessible $(T,U)$ values and, consequently, the location of the consensus--fragmentation transition. We examine this effect in the following section.

To characterize the dynamics underlying the stationary states, we next examine the time evolution of agents' beliefs for four representative points, A--D, indicated in Fig.~\ref{FullyRes}(c). Figures~\ref{FullyRes}(d) and (e) show two parameter sets within the consensus region. In both cases, all agents eventually converge to the same belief, $b=0$, corresponding to a moderate consensus. Although the final state is the same, the convergence time depends strongly on the threshold values. For the parameter set with low $T_T$ and high $U_T$ (point A), the system rapidly approaches the stationary state. As the thresholds approach the consensus--fragmentation boundary (point B), the convergence becomes substantially slower, as reflected by the different time scales in Figs.~\ref{FullyRes}(d) and (e). Thus, the proximity to the transition boundary strongly affects the timescale over which consensus is established.

The dynamics in the fragmentation region are qualitatively different. As illustrated in Figs.~\ref{FullyRes}(f) and (g), parameter sets close to the transition boundary can lead to the formation of local consensus groups, in which subsets of agents converge to similar belief values while distinct groups remain separated. This process reduces the overall diversity of beliefs relative to the initial state. For the parameter set corresponding to point D, however, the dynamics become frozen: the agents' initial beliefs remain essentially unchanged throughout the simulation. The resulting fragmented state therefore retains the initial diversity of opinions, in contrast to the partial convergence observed for the other fragmented case.

These distinct dynamical behaviors---rapid or slow convergence to consensus, partial convergence into local consensus groups, and frozen fragmentation---motivate a more detailed analysis of the mechanisms governing the stationary states and their approach to equilibrium, which we consider in Sec.~\ref{secAnalysis}.

\subsubsection{Correlated trust and distrust}

\begin{figure*}[t!]
	\includegraphics[scale=0.5]{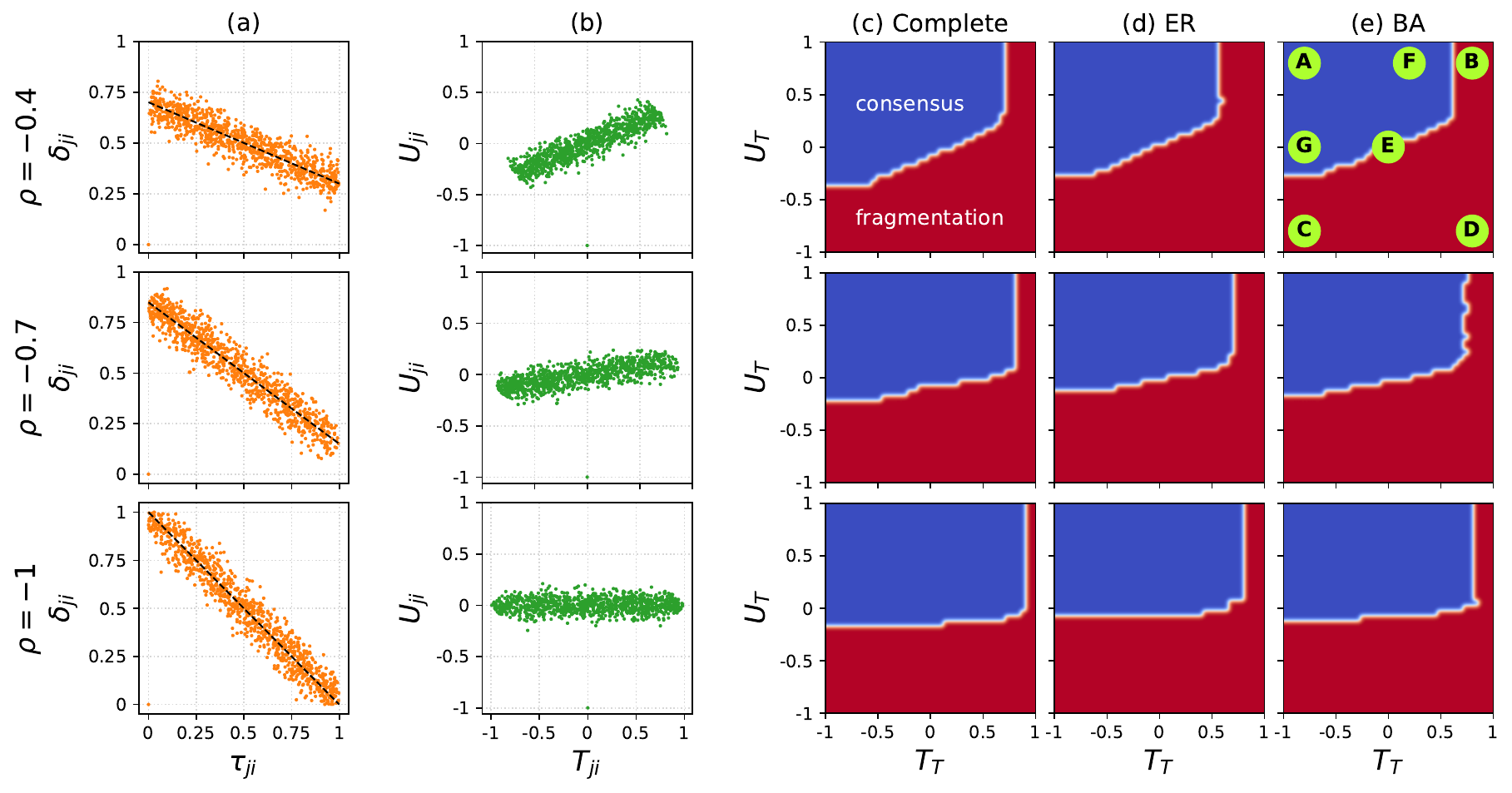}
	\centering
	\caption{\textbf{Phase diagrams for correlated trust and distrust degrees with $\sigma_{\epsilon}=0.05$.} (a) Correlated trust-degree $\tau_{ji}$ and distrust-degree $\delta_{ji}$ values, and (b) the corresponding net trust $T_{ji}$ and uncertainty $U_{ji}$ values for agent $i$ toward its neighbors $j$. (c)--(e) Phase diagrams in the $(T_T,U_T)$ parameter space for (c) the complete graph, (d) the Erdős--Rényi random network, and (e) the Barabási--Albert network. For the latter two network structures, $N=1000$ and $\langle k\rangle=100$. From top to bottom, the magnitude of the negative correlation between $\tau_{ji}$ and $\delta_{ji}$ increases. The boundaries of consensus (blue) and fragmentation (red) regions are estimated from a single realizations; they may vary due to finite-sample variability.}
	
	\label{phase05}
\end{figure*}

\begin{figure*}[t!]
	\includegraphics[scale=0.5]{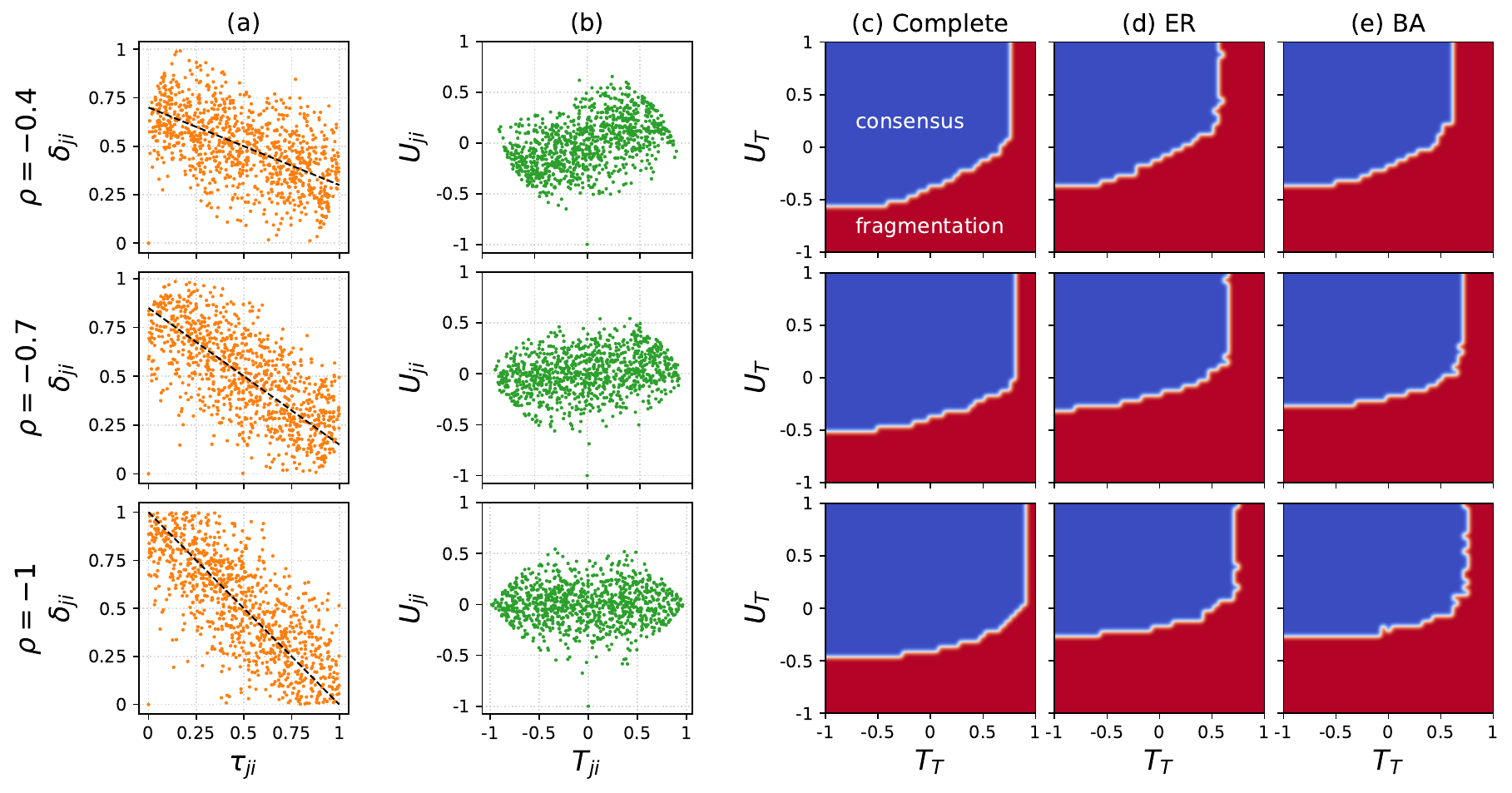}
	\centering
	\caption{\textbf{Phase diagrams for correlated trust and distrust degrees with $\sigma_{\epsilon}=0.15$.} (a) Correlated trust-degree $\tau_{ji}$ and distrust-degree $\delta_{ji}$ values, and (b) the corresponding net trust $T_{ji}$ and uncertainty $U_{ji}$ values for agent $i$ toward its neighbors $j$. (c)--(e) Phase diagrams in the $(T_T,U_T)$ parameter space for (c) the complete graph, (d) the Erdős--Rényi random network, and (e) the Barabási--Albert network. For the latter two network structures, $N=1000$ and $\langle k\rangle=100$. From top to bottom, the magnitude of the negative correlation between $\tau_{ji}$ and $\delta_{ji}$ increases. The boundaries of consensus (blue) and fragmentation (red) regions are estimated from a single realizations; they may vary due to finite-sample variability.}
	\label{phase15}
\end{figure*}

Trust and distrust degrees are generally expected to be negatively correlated. To incorporate this dependence, we generate the initial values of $\tau$ and $\delta$ using the procedure described in Sec.~\ref{secModel_01}. Figures~\ref{phase05}(a) and \ref{phase15}(a) show the resulting trust degree $\tau_{ji}$ and distrust degree $\delta_{ji}$ assigned by agent $i$ to its neighbors for $\sigma_{\epsilon}=0.05$ and $0.15$, respectively. In each figure, the magnitude of the negative correlation increases from top to bottom. Compared with the uncorrelated case shown in Fig.~\ref{FullyRes}(a), the $(\tau,\delta)$ values are no longer distributed throughout the entire unit square. Instead, they become increasingly concentrated around the relation $\delta_{ji}=\rho(\tau_{ji}-0.5)+0.5,
$ as the magnitude of the negative correlation increases. Thus, the correlation between trust and distrust degrees restricts the combinations of trust degree and distrust degree that can occur across links.

Applying Eqs.~(\ref{TEqu})--(\ref{UEqu}) yields the corresponding net trust $T_{ji}$ and uncertainty degree $U_{ji}$ shown in Figs.~\ref{phase05}(b) and \ref{phase15}(b). The $(T,U)$ pairs likewise occupy a more restricted region than in the uncorrelated case, where they are distributed over a broader diamond-shaped region (Fig.~\ref{FullyRes}(b)). Consequently, for a given pair of trust and uncertainty thresholds $(T_T,U_T)$, the number of neighbors satisfying the conditions required to become influential neighbors can be reduced. This reduction in the availability of INs is reflected in a smaller consensus region in the $(T_T,U_T)$ phase diagram.

For the complete graph, this effect is evident in Figs.~\ref{phase05}(c) and \ref{phase15}(c). Compared with the uncorrelated case [Fig.~\ref{FullyRes}(c)], the consensus region becomes smaller as the magnitude of the negative correlation increases. For a fixed $\sigma_{\epsilon}$, the phase-transition boundary also shifts toward higher values of the trust and uncertainty thresholds. In addition, the steep part of the boundary becomes less pronounced, indicating a more gradual transition between the consensus and fragmentation regimes. These changes are more clearly visible for $\sigma_{\epsilon}=0.05$ (Fig.~\ref{phase05}(c)) than for $\sigma_{\epsilon}=0.15$ (Fig.~\ref{phase15}(c)), where the broader distribution of $(T,U)$ values reduces the effect of the correlation on the phase diagram.

%%%%%%%%%%%%%%%%%%%%%%%%%%%
%%%%%%%%%%%%%%%%%%%%%%%%%%%
%%%%%%%%%%%%%%%%%%%%%%%%%%%

\subsection{Homogeneous networked population} \label{secResults_02}

We next consider networked populations in which each agent interacts with only a subset of the population. We first examine a homogeneous network structure, in which agents have approximately the same number of neighbors. Specifically, we consider an Erdős--Rényi random network of size $N=1000$ with mean degree $\langle k\rangle=100$. The resulting phase diagrams for correlated trust and distrust degrees are shown in Figs.~\ref{phase05}(d) and \ref{phase15}(d) for $\sigma_{\epsilon}=0.05$ and $0.15$, respectively.

Compared with the corresponding complete-graph cases (Figs.~\ref{phase05}(c) and \ref{phase15}(c)), the consensus region is substantially reduced. This reduction results from the smaller number of available neighbors: with fewer neighbors, an agent has fewer opportunities to find neighbors that satisfy the conditions for becoming influential neighbors. The effect is particularly pronounced for $\sigma_{\epsilon}=0.05$, for which the correlated $(T,U)$ values are more concentrated. In this case, the restricted set of possible net trust and uncertainty-degree combinations further reduces the availability of INs and leads to a stronger shift of the consensus--fragmentation boundary.

The dependence on the trust--distrust correlation observed for the complete graph is also retained in the Erdős--Rényi network. For both values of $\sigma_{\epsilon}$, increasing the magnitude of the negative correlation modifies the location and shape of the consensus--fragmentation boundary. Thus, the emergence of consensus depends jointly on the statistical structure of trust and distrust degrees and on the connectivity of the underlying social network.

\subsection{Heterogeneous networked population}

We next consider a network structure with a heterogeneous number of neighbors. To this end, we perform simulations on Barabási--Albert (BA) networks with the same network size and mean degree as the Erdős--Rényi networks considered above. Figures~\ref{phase05}(e) and \ref{phase15}(e) show the corresponding phase diagrams for $\sigma_{\epsilon}=0.05$ and $0.15$, respectively.

Interestingly, the phase diagrams of the BA networks are very similar to those of the corresponding Erdős--Rényi networks, as can be seen by comparing panels (d) and (e) of Figs.~\ref{phase05} and \ref{phase15}. Thus, despite the presence of highly connected hubs and the resulting heterogeneity in the degree distribution, the equilibrium consensus--fragmentation boundary is not substantially altered compared with the homogeneous network structure. This result suggests that degree heterogeneity has a more limited effect on the location of the consensus--fragmentation transition than on the specific equilibrium state that emerges when hubs are assigned biased initial beliefs. The qualitative robustness of these phase diagrams to a larger network population is further examined in Appendix~\ref{App0}, where we consider $N=5000$ with $\langle k^{out}\rangle=50$ for both ER and BA networks. The overall structure of the consensus--fragmentation boundaries remains similar, although some quantitative shifts are observed.

This observation raises the question of whether hubs may instead influence the dynamics leading to the equilibrium state, even when the final phase diagram remains largely unchanged. It is important to note that the initial beliefs of all agents are drawn uniformly from the interval $[-5,5]$. Moreover, although the trust and distrust degrees are negatively correlated, their values are assigned randomly across neighboring agents and are therefore not explicitly associated with network degree. Consequently, the present setup does not introduce any systematic relationship between an agent's initial belief, its trust and distrust degrees, and its network degree. To investigate whether hubs can play a more specific role under different conditions, we therefore consider additional scenarios in which the heterogeneity of agents' degrees is explicitly taken into account.

\subsubsection{Biased trust and distrust} \label{bisedtandd}

We first examine a scenario in which agents with higher degrees receive greater trust degree and lower distrust degree from their neighbors, following the procedure described in Sec.~\ref{secModel_02}. To assess the effect of this degree-dependent bias, we perform simulations for two different swapping thresholds, $p_s=0.5$ and $p_s=1$. The resulting phase-transition boundary in the $(T_T,U_T)$ parameter space remains essentially unchanged from the unbiased case, $p_s=0$ (results not shown).

This result can be understood from the way the bias is introduced into the model. Swapping the trust degree and distrust degree assigned to neighbors does not change the set of $(\tau_{ji},\delta_{ji})$ values; it only changes which neighbors are associated with particular values in this space. Consequently, under the swapping procedure, higher-degree neighbors are preferentially assigned values located in the lower-right region of the $(\tau,\delta)$ space in Figs.~\ref{phase05}(a) and \ref{phase15}(a). Through Eqs.~(\ref{TEqu})--(\ref{UEqu}), these values correspond to the upper-right region of the $(T,U)$ space shown in Figs.~\ref{phase05}(b) and \ref{phase15}(b) (see Appendix~\ref{AppA}), where $T$ and $U$ denote net trust and uncertainty degree, respectively. Thus, the degree-dependent bias changes which neighbors are more likely to satisfy the conditions for becoming influential neighbors for a given pair of trust and uncertainty thresholds $(T_T,U_T)$. In particular, depending on the threshold values, the INs can contain a larger or smaller fraction of high-degree neighbors, including network hubs.

However, the degree of an IN does not directly enter the belief-update rule in Eq.~(\ref{andEqu}). Once a neighbor satisfies the conditions for being an IN, its contribution to the evolution of an agent's belief depends on the neighbor's belief $b_j$, rather than on its degree. Therefore, the degree-dependent reassignment of trust and distrust degrees changes the identity of the INs but does not otherwise modify their contribution to belief propagation.

Given that the initial beliefs are drawn uniformly from $[-5,5]$ and are not correlated with network degree, whether the selected INs are hubs or low-degree neighbors does not systematically alter the distribution of their beliefs. Consequently, changing the swapping threshold $p_s$ does not lead to a substantial change in the equilibrium phase diagram. This result indicates that, under the present model assumptions, degree-dependent reassignment of trust and distrust degrees alone is insufficient to substantially modify the consensus--fragmentation transition.	

%%%%%%%%%%%%%%%%%%%%%%%%%%%	
\begin{figure*}[t!]
	\includegraphics[scale=0.55]{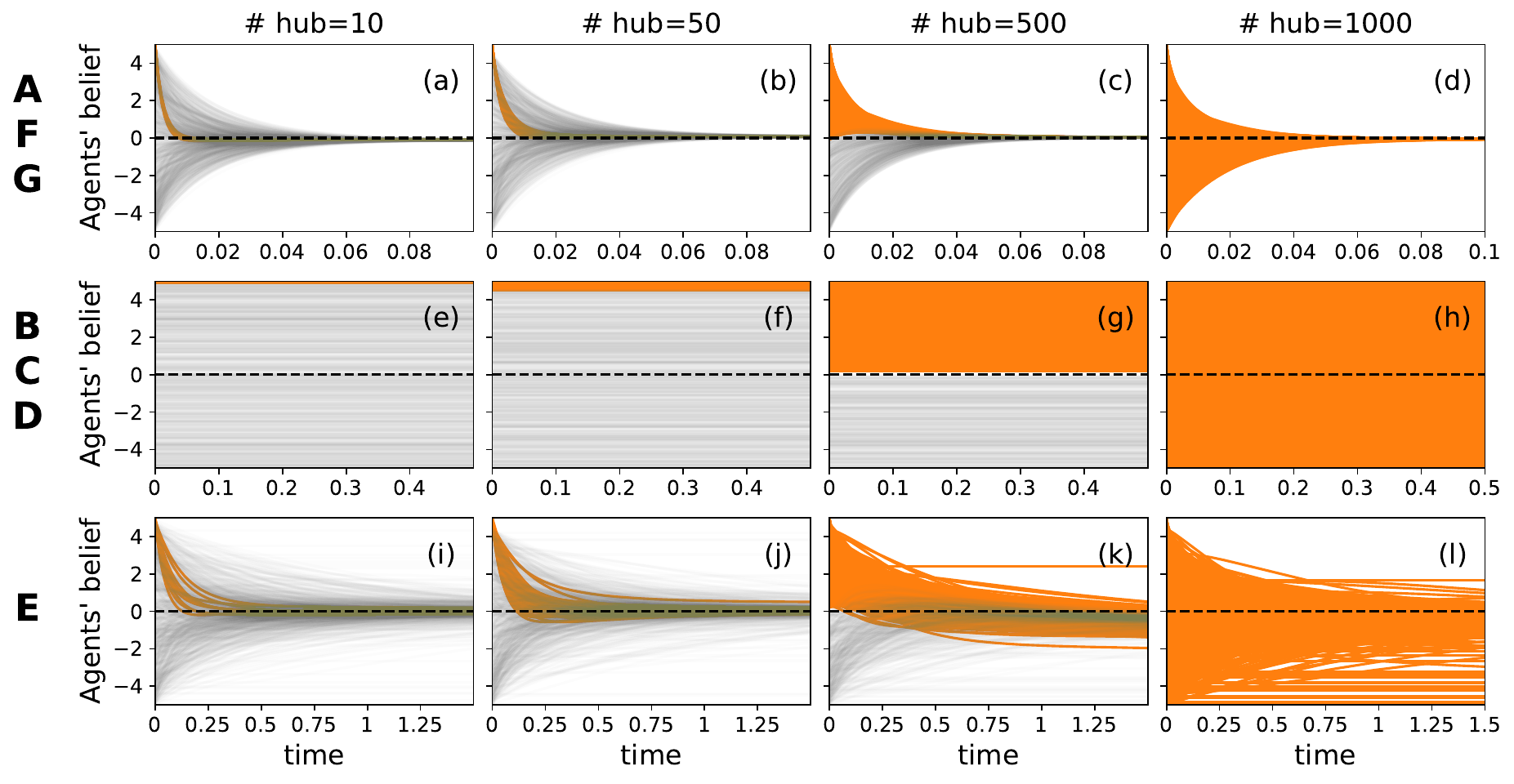}
	\centering
	\caption{\textbf{Time evolution of agents' beliefs for different numbers of radical hubs with $p_s=0$.} The trust and contradiction thresholds for each row correspond to the representative points shown in Fig.~\ref{phase05}(e). Orange curves represent the beliefs of radical hubs, while gray curves represent the beliefs of the other agents. The number of radical hubs increases from left to right. Other parameters are the same as those used in Fig.~\ref{phase05}(e).}
	\label{btimes0}
\end{figure*}
%%%%%%%%%%%%%%%%%%%%%%%%%%%	
\begin{figure*}[t!]
	\includegraphics[scale=0.55]{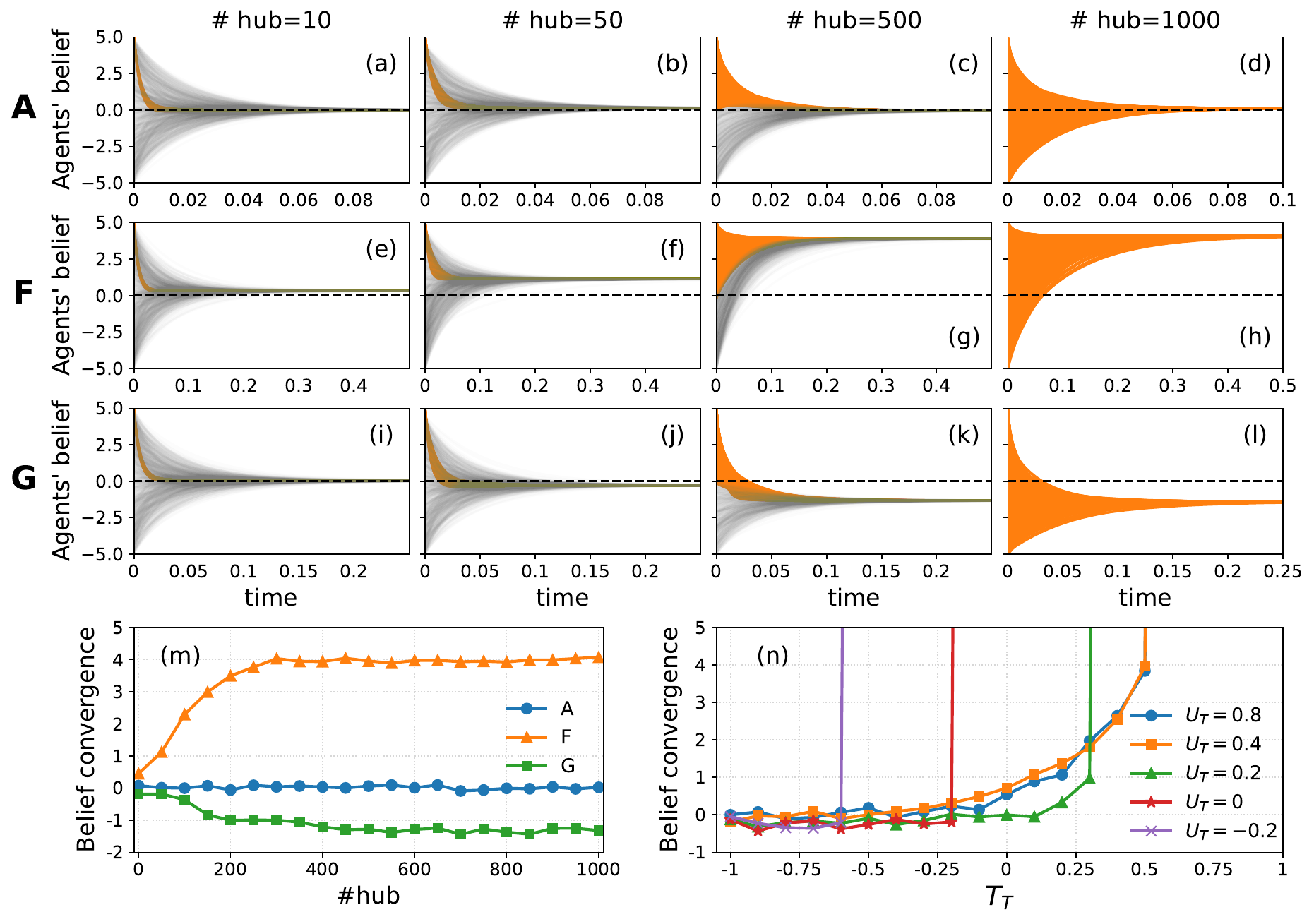}
	\centering
	\caption{\textbf{Time evolution of agents' beliefs for different numbers of radical hubs with $p_s=1$.} The trust and contradiction thresholds for each row correspond to the representative points shown in Fig.~\ref{phase05}(e), with $(T_A,U_A)=(-0.8,0.8)$, $(T_F,U_F)=(0.2,0.8)$, and $(T_G,U_G)=(-0.8,0)$. The behavior for cases B, C, D, and E is similar to that shown in Fig.~\ref{btimes0}. Orange curves represent the beliefs of radical hubs, while gray curves represent the beliefs of the other agents. The number of radical hubs increases from left to right. Other parameters are the same as those used in Fig.~\ref{phase05}(e). (m) Convergence belief (CB) as a function of the number of radical hubs $h$ for cases A, F, and G. (n) CB as a function of the trust thresholds $T_T$ for different uncertainty threshold $U_T$ at $h=50$. The vertical lines indicate the maximum value of $T_T$ for which consensus can be reached.}
	\label{btimes1}
\end{figure*}
%%%%%%%%%%%%%%%%%%%%%%%%%%%

\subsubsection{Biased beliefs}

The results presented above show that assigning greater trust degree and lower distrust degree to higher-degree agents does not substantially modify the equilibrium phase-transition boundary. Although the degree of an influential neighbor does not explicitly enter the belief-update rule, network degree can still affect the dynamics indirectly through the number of neighbors available to an agent. This raises the question of whether assigning initially biased beliefs to highly connected agents can alter the equilibrium state, even when the phase diagram itself remains unchanged.

To address this question, we consider a heterogeneous population in which $h$ hubs are assigned the radical initial belief $b=+5$, while the remaining agents retain uniformly distributed initial beliefs, as described in Sec.~\ref{secModel_02}. We first consider the unbiased assignment of trust and distrust degrees, $p_s=0$. Varying the number of radical hubs produces no detectable change in the phase diagram, regardless of the value of $h$ (results not shown). We also repeat the simulations for $p_s=1$, thereby introducing a degree-dependent bias in both the initial beliefs and the trust degree and distrust degree assignments. In this case, the phase-transition boundary likewise remains unchanged from that of the BA network without biased quantities. Thus, the presence of radical hubs affects the trajectory and final consensus belief within the consensus regime, rather than the location of the consensus--fragmentation transition itself.

To understand this result and the subsequent effects of biased beliefs on the dynamics, we first examine the temporal evolution of the agents' beliefs for $p_s=0$. Figure~\ref{btimes0} shows the time evolution for several combinations of trust and uncertainty thresholds corresponding to the representative points in Fig.~\ref{phase05}(e). Figures~\ref{btimes0}(a)--(d) correspond to parameter sets A, F, and G within the consensus region. In all three cases, the system eventually reaches consensus, and the stationary belief remains close to the moderate value $b=0$, even as the number of radical hubs increases from left to right.

For $p_s=0$, the correlated trust and distrust degrees are assigned randomly to the neighbors, independently of their network degree. Consequently, for a given pair of trust and uncertainty thresholds $(T_T,U_T)$, a hub and a low-degree neighbor have the same probability of being classified as an IN on a per-neighbor basis. Increasing $h$, however, creates a strong correlation between network degree and initial belief: highly connected agents are assigned beliefs close to $b=+5$, whereas low-degree agents are predominantly associated with beliefs closer to $b=-5$. The hubs and low-degree agents therefore represent two groups with initially opposing beliefs.

Although the probability of any individual neighbor being an IN is independent of degree, hubs have substantially more neighbors and consequently tend to have a larger number of INs. Their beliefs are therefore updated more frequently than those of low-degree agents. Moreover, because their INs include agents with a broad range of initial beliefs, the repeated updates drive the beliefs of the hubs rapidly toward more moderate values. This effect is visible in the steeper temporal trajectories of the hubs in Fig.~\ref{btimes0}(a). In contrast, low-degree agents have fewer INs and consequently update their beliefs more slowly. The resulting difference in update rates progressively reduces the initial separation between high- and low-degree agents and drives the population toward a moderate consensus, as observed in Figs.~\ref{btimes0}(a)--(d).

The situation is different in the fragmentation regime. As shown in Figs.~\ref{btimes0}(e)--(h), for the parameter sets corresponding to points B, C, and D, the system can enter a frozen state in which the agents' beliefs remain essentially unchanged. In this regime, the absence of INs prevents belief updates, so changing the number of radical hubs has no effect on the temporal evolution. Importantly, assigning a biased initial belief to the hubs does not itself change the number of INs. Similarly, for the parameter set corresponding to point E, the system reaches a fragmented stationary state even when the number of radical hubs is large, as shown in Figs.~\ref{btimes0}(i)--(l). Thus, for $p_s=0$, the initial bias of hubs does not alter the conditions determining whether the system reaches consensus or fragmentation.

We next introduce a degree-dependent bias in the trust degree and distrust degree assignments by setting $p_s=1$. The corresponding temporal dynamics are shown in Fig.~\ref{btimes1}, where each row represents a different set of trust and uncertainty thresholds associated with the points in Fig.~\ref{phase05}(e). For point A, characterized by a very low trust threshold and a high uncertainty threshold, the system again approaches a moderate consensus, similarly to the $p_s=0$ case [Fig.~\ref{btimes1}(a)--(d)]. The fragmentation behavior observed for points B--E is also qualitatively similar to that shown in Fig.~\ref{btimes0} (results for these cases are not shown).

A qualitatively different behavior emerges for point F, which lies in the consensus region but has a higher trust threshold than point A. As shown in Fig.~\ref{btimes1}(e)--(h), the stationary consensus becomes positively biased, particularly as the number of radical hubs increases. The larger the number of hubs initially assigned $b=+5$, the stronger the positive bias of the final consensus. Thus, although the phase diagram is unchanged, the final consensus belief within the consensus regime can depend strongly on the initial beliefs of highly connected agents.

Interestingly, the direction of the bias can be reversed for point G. As shown in Fig.~\ref{btimes1}(i)--(l), the final consensus can become negatively biased when the contradiction threshold is lower, despite assigning the more positive initial bias to the hubs. Thus, the positive bias of highly connected
agents does not necessarily determine the direction of the final consensus. Instead, the trust and distrust thresholds can preferentially select lower-degree agents as influential neighbors, allowing their initially negative-biased beliefs to dominate the propagated influence.

To quantify these effects, Fig.~\ref{btimes1}(m) shows the convergence belief (CB) as a function of the number of radical hubs $h$ for points A, F, and G. For cases F and G, the CB approaches its final value at approximately $h\approx400$, beyond which further increasing the number of radical hubs produces little additional change. Notably, the CB for case G, illustrated by the green squares, remains negative even as the number of positively biased hubs is increased. This demonstrates that increasing the number of
positively biased hubs does not necessarily reverse the direction of the consensus when the threshold conditions preferentially select
lower-degree agents as INs. Figure~\ref{btimes1}(n) shows the dependence of the CB on the trust thresholds for different uncertainty threshold. Within the consensus regime, increasing $T_T$ shifts the CB toward higher values until the critical threshold at which fragmentation occurs.

The contrasting behaviors observed for cases F and G can be understood from the joint effect of the degree-dependent trust degree and distrust degree assignment and the trust and uncertainty thresholds. When $p_s=1$, higher-degree agents are preferentially associated with greater trust degree and lower distrust degree, as discussed in Sec.~\ref{bisedtandd}. In the $(T,U)$ space, this preferential assignment places high-degree neighbors toward the upper-right region, whereas lower-degree neighbors are preferentially associated with the lower-left region. The trust and uncertainty thresholds $(T_T,U_T)$ therefore determine not only the total number of INs but also their degree composition.

In particular, we selected cases F and G with $(T_F,U_F)=(0.2,0.8)$ and $(T_G,U_G)=(-0.8,0)$, respectively. Although these two points have different threshold values, they yield approximately similar numbers of INs according to Eq.~(\ref{kappaEqu}). Their degree composition, however, is different. The higher values of both thresholds in case F increase the relative likelihood that high-degree agents are selected as INs. Because the hubs are initially assigned the positive belief $b=+5$, the beliefs of their neighbors are therefore more strongly driven toward positive values. In contrast, the lower thresholds in case G favor the selection of a larger fraction of low-degree neighbors as INs. Since these agents are predominantly associated with the negative side of the initial belief distribution, their influence can shift the final consensus toward negative values.

Increasing $h$ further strengthens this effect because it increases the fraction of INs carrying the biased belief. In case F, the distribution of IN beliefs therefore becomes increasingly weighted toward $b=+5$, producing a progressively more positive consensus. Conversely, in case G, the selected INs can be increasingly dominated by agents with beliefs near $b=-5$, resulting in a progressively more negative consensus. Once the relevant IN population is effectively dominated by the biased hubs, however, adding further biased hubs does not substantially change its belief composition. This provides an explanation for the saturation of the CB observed for $h\gtrsim400$ in Fig.~\ref{btimes1}(m).

Overall, these results reveal an important distinction between the equilibrium phase boundary and the final consensus belief within the consensus regime. For the heterogeneous networks considered here, introducing biased initial beliefs does not substantially modify the consensus--fragmentation transition. Nevertheless, when trust degree and distrust degree are also biased according to network degree, the combination of the degree distribution, the trust and uncertainty thresholds $T_T$ and $U_T$, and the number of radical hubs $h$ can strongly affect which agents become INs and, consequently, the value and direction of the final consensus. In contrast, when the hubs are assigned moderate initial beliefs, we find that the final consensus remains moderate over the range of $T_T$, $U_T$, and $p_s$ examined (results not shown).

%% file: 050_analysis.tex
\section{\label{secAnalysis}Comfort index analysis}
%===========================
%%%%%%%%%%%%%%%%%%%%%%%%%%%	
\begin{figure*}[t!]
	\includegraphics[scale=0.65]{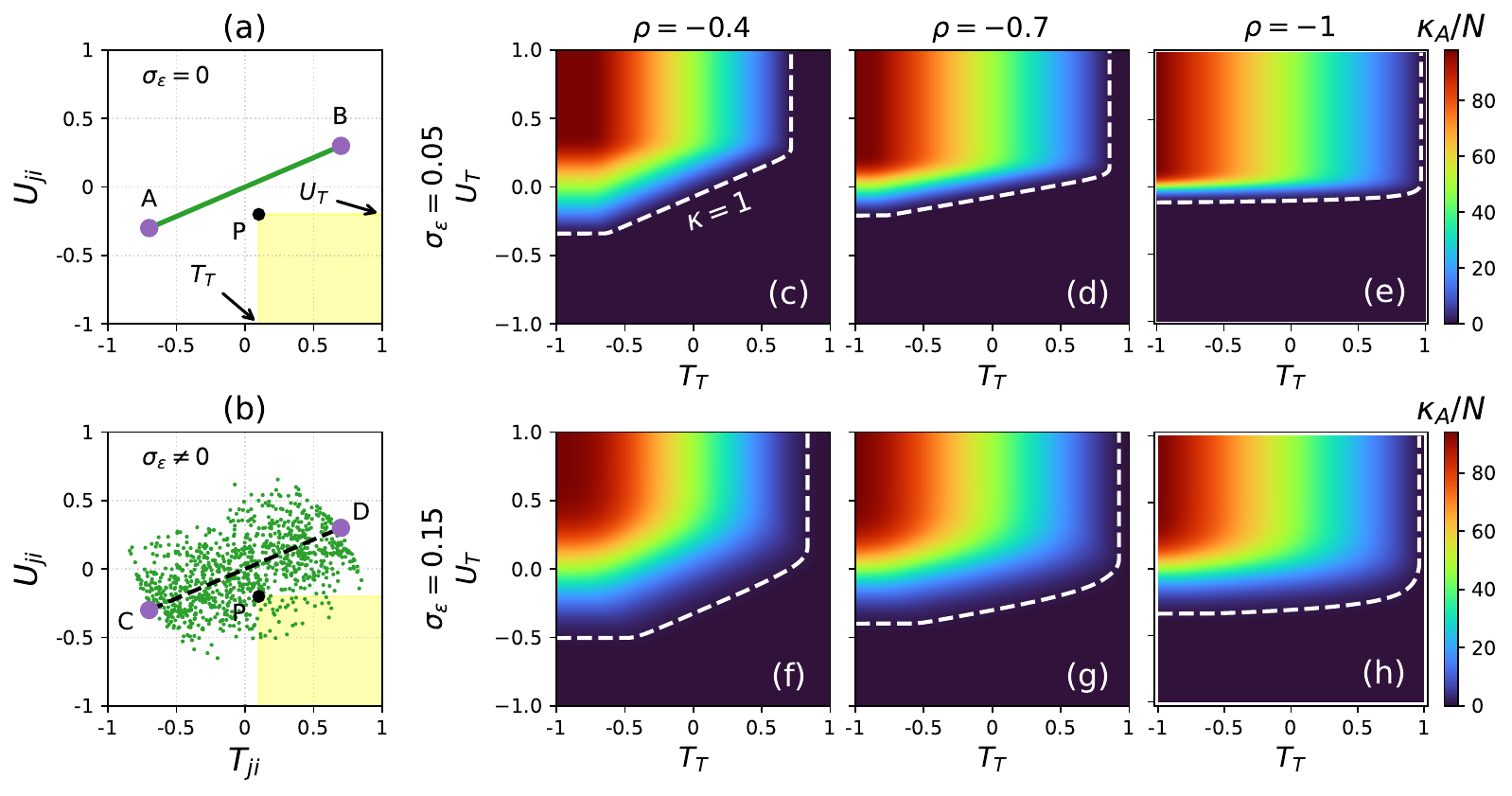}
	\centering
	\caption{\textbf{Geometric interpretation of influential neighbors and
			comfort-index landscapes.}
		Green dots represent the trust and uncertainty degrees of agent $i$ with
		respect to its neighbors $j$ for $\rho=-0.4$ in (a) the noise-free case,
		$\sigma_\epsilon=0$, and (b) the finite-noise case, $\sigma_\epsilon\neq0$.
		For the representative threshold pair $(T_T,U_T)$ indicated by point $P$,
		the yellow shaded region denotes the interaction territory defined by
		$T>T_T$ and $U<U_T$. In (a), the admissible $(T,U)$ values lie on the
		deterministic line connecting $A$ and $B$, whereas in (b) finite noise
		broadens their distribution around this relation. Panels (c)--(h) show
		the aggregate comfort index per agent,  $\frac{\kappa_A}{N}$, obtained by numerically evaluating
		Eq.~(\ref{FequFinal}) for $\langle k^{out}\rangle=100$. Panels (c)--(e)
		correspond to $\sigma_\epsilon=0.05$ and panels (f)--(h) to
		$\sigma_\epsilon=0.15$. From left to right, the magnitude of the negative
		correlation increases, with $\rho=-0.4$, $-0.7$, and $-1$. The white dashed
		curves indicate the contour $\kappa=1$, corresponding to one expected
		influential neighbor per agent.}
	\label{comfortIndex1Fig}
\end{figure*}
%%%%%%%%%%%%%%%%%%%%%%%%%%%

The results presented above show that the availability of influential neighbors that satisfy the two influence conditions plays a central role in determining whether belief propagation can occur and in shaping the transition between consensus and fragmentation. This observation can be visualized
geometrically in the $T-U$ plane, where the thresholds $T_T$ and $U_T$ define
an \textit{interaction territory} (IT) given by
$T>T_T$ and $U<U_T$. The IT therefore identifies the region of the
$T-U$ plane in which a neighbor can influence the belief of agent $i$.
Figure~\ref{comfortIndex1Fig}(a)--(b) illustrates this construction for
$\rho=-0.4$. The green points represent the trust and uncertainty degrees
associated with the neighbors of a given agent, while the yellow region
corresponds to the IT for the representative thresholds indicated by point
$P$.

When the IT overlaps substantially with the region occupied by the
$(T_{ji},U_{ji})$ values, a relatively large number of neighbors can satisfy
the influence conditions. Consequently, agents have more opportunities to
update their beliefs, which generally facilitates the propagation of
information through the network. This geometric picture also provides an
interpretation of the different convergence times observed in the consensus
regime of Fig.~\ref{FullyRes}: parameter sets for which more neighbors are
available as INs generally lead to faster belief evolution, whereas
approaching the consensus--fragmentation boundary reduces the availability
of INs and can substantially slow down convergence.

The role of the distribution of trust and distrust becomes particularly clear
when comparing Fig.~\ref{comfortIndex1Fig}(a) and (b). In the absence of
noise, the $(T,U)$ values are confined to the deterministic line derived in
Appendix~\ref{AppA}. Consequently, whether the IT contains any admissible
values is determined entirely by the position of the threshold pair
$(T_T,U_T)$ relative to this line. If the IT does not intersect the admissible
set, no IN can exist and the agents' beliefs cannot change.
The corresponding geometric boundary is therefore a natural criterion for
the loss of belief propagation. In the noise-free limit, this boundary provides a necessary condition for the existence of influential neighbors; for the complete graph in the thermodynamic limit, it is both necessary and sufficient for INs to exist. 

For finite noise, the deterministic line broadens into a distribution in the
$T-U$ plane, as illustrated in Fig.~\ref{comfortIndex1Fig}(b). As a result, the IT can contain influential neighbors even when
the corresponding threshold pair would not intersect the noise-free line.
Increasing $\sigma_\epsilon$ therefore broadens the range of threshold
values for which INs are available. This provides a geometric interpretation of the expansion of the consensus region observed when comparing the phase diagrams for $\sigma_\epsilon=0.05$ and
$\sigma_\epsilon=0.15$ in Figs.~\ref{phase05} and~\ref{phase15}. The effect
of the correlation parameter is also apparent in the $T-U$ representation:
changing $\rho$ modifies the orientation of the distribution of trust and
uncertainty degrees and therefore changes its overlap with the IT.

To quantify the availability of influential neighbors in the presence of
noise, we consider the fraction $F$ introduced in Eq.~(\ref{Fequ}). As
derived in Appendix~\ref{AppB}, the joint density of $T$ and $U$ allows $F$
to be obtained by integrating over the interaction territory. Thus,
$F(T_T,U_T,\rho,\sigma_\epsilon)$ represents the expected fraction of
outgoing neighbors that satisfy both influence conditions under the analytical approximation described in Appendix~\ref{AppB}. For an agent with out-degree $k_i^{out}$, this fraction is converted into
the comfort index through Eq.~(\ref{kappaEqu}), such that $\kappa_i$
represents the expected number of INs of that agent.
Thus, $F$ characterizes the availability of influential relationships as a
fraction of the outgoing neighborhood, whereas $\kappa_i$ additionally
accounts for the agent's connectivity (see Eq.~\ref{FequFinal}). For a population-level characterization, however, the aggregate comfort index
introduced in Sec.~\ref{secModel} is
\begin{equation}
	\kappa_A
	= \sum_{k=k_{\min}}^{k_{\max}} N P_k \kappa_k
	= N F(T_T,U_T,\rho,\sigma_\epsilon)
	\langle k^{out}\rangle .
	\label{kappaAggregateAnalysis}
\end{equation}
Hence, $\kappa_A$ is the expected total number of influential connections
in the network. Equivalently, $\kappa_A/N =
F\langle k^{out}\rangle$ gives the expected number of INs
per agent at the population level. For a fixed mean degree, the aggregate
quantity therefore has the same dependence on $(T_T,U_T,\rho,\sigma_\epsilon)$ as the individual comfort-index landscapes, differing only by a scale
factor. The resulting population-level comfort landscapes are shown in Fig.~\ref{comfortIndex1Fig}(c)--(h). For $\langle k^{out}\rangle=100$, this quantity is $\kappa_A/N=100F$. In all panels, $\kappa_A/N$ increases as the trust threshold is relaxed and the
uncertainty threshold is increased, reflecting the larger interaction
territory available to the agents. Regions with large $\kappa_A/N$ therefore
correspond to parameter ranges in which a substantial fraction of the
outgoing neighbors are expected to satisfy the influence conditions, whereas
small values of $\kappa_A/N$ indicate that influential neighbors are unlikely to
be available.

The dependence on $\rho$ and $\sigma_\epsilon$ is also visible in the shape
of aggregate comfort index  landscapes. Increasing the magnitude of the negative
correlation changes the orientation of the underlying distribution in the
$T-U$ plane and consequently modifies the threshold combinations required
to obtain a given level of IN availability. Increasing
$\sigma_\epsilon$, in contrast, broadens the distribution around the
noise-free relation. This broadening allows a nonzero fraction of neighbors
to enter the IT over a wider range of threshold values and makes the
transition in $\kappa_A/N$ more gradual. Thus, the changes observed in the
aggregate comfort-index landscapes provide a geometric counterpart to the shifts and
broadening of the consensus--fragmentation boundaries obtained directly
from simulations.

The white dashed curves indicate $\kappa_A/N=1$, corresponding to one expected influential connection per agent at the population level. This contour
should not be interpreted as an exact consensus--fragmentation boundary.
Rather, it provides a useful reference scale: for $\frac{\kappa_A}{N} \ll1$, an agent
is likely to have no influential neighbor, whereas values of $\frac{\kappa_A}{N}$ of
order unity indicate that INs become available on
average. Whether the entire network reaches consensus depends on the
collective connectivity and arrangement of these influential relationships,
and therefore cannot be determined from the individual-level value
$\kappa_i$ alone. The close correspondence between these analytical landscapes and
the consensus--fragmentation patterns obtained from direct simulations supports the use of the comfort index as a compact measure of the
availability of influential relationships underlying the observed collective dynamics.

%% file: 060_discussion.tex
\section{\label{secDiscussion}Discussion }
%===========================
In this work, we investigated how the joint structure of trust and distrust relationships affects belief propagation and the emergence of consensus or fragmentation in a networked population. Rather than treating trust and distrust as opposite states of a single relation, we considered them as two distinct but coupled quantities \cite{marsh2005trust,kawabe2019trust,lewicki1998trust} and characterized each relationship through its net trust and uncertainty degree. Under this representation, a neighbor can influence an agent only when both the trust and uncertainty conditions are satisfied. This provides a direct link between the statistical properties of trust and distrust and the availability of influential neighbors (INs), which is closely related to whether belief propagation can be sustained throughout the network \cite{ziegler2005propagation}. Our results show that this link provides a useful way of understanding both the consensus--fragmentation transition and the final consensus belief.

A central result is that the correlation between trust and distrust can substantially alter the collective dynamics. Negative correlation restricts the set of admissible trust--distrust combinations and consequently changes the region occupied by the corresponding net trust and uncertainty degrees. As a result, the same values of the trust and contradiction thresholds can give rise to different numbers of INs depending on the correlation structure of the underlying relationships. Increasing the magnitude of the negative correlation therefore shifts and reshapes the consensus--fragmentation boundary. This observation emphasizes that the marginal levels of trust or distrust alone are insufficient to characterize the propagation of influence: their joint organization determines which relationships satisfy the two influence conditions \cite{marsh2005trust, lewicki1998trust, zhang2024profiling}. In this sense, the correlation between trust and distrust acts as a structural constraint on the set of relationships through which information can propagate.

The network results further distinguish the role of connectivity from that of degree heterogeneity. Compared with the complete network, the Erd\H{o}s--R\'enyi networks exhibit a reduced consensus region, consistent with the smaller number of available neighbors and hence fewer opportunities for influential relationships. In contrast, for networks with comparable size and mean degree, the equilibrium phase diagrams of Erd\H{o}s--R\'enyi and Barab\'asi--Albert networks remain qualitatively similar. Thus, connectivity has a substantial effect on the availability of influence, whereas degree heterogeneity 
alone has a more limited effect on the location of the consensus--fragmentation transition under the present update rule. The larger-network results in Appendix~\ref{App0} provide further evidence that the overall structure of the phase diagrams is robust when the network population is increased, although quantitative shifts in the phase boundaries are observed. Importantly, these calculations change both the network size and the mean degree and therefore do not constitute an isolated finite-size analysis.

The biased-belief experiments reveal another aspect of the dynamics that is not captured by the equilibrium phase boundary alone. Assigning extreme initial beliefs to highly connected agents does not substantially change the location of the consensus--fragmentation transition. Nevertheless, it can strongly affect the trajectory and the value of the final consensus within the consensus regime. This distinction arises because the number of INs and the identity of those influential neighbors are not equivalent. When trust and distrust are assigned independently of degree, high-degree agents are not preferentially selected as INs, and assigning them extreme initial beliefs mainly affects their own trajectories. When trust and distrust are instead correlated with degree, the threshold conditions can preferentially select different parts of the degree distribution as INs. Consequently, the beliefs carried by those selected agents can determine the direction and magnitude of the final consensus \cite{mengers2024leveraging}.

The comparison between cases F and G in Fig.~\ref{btimes1} provides a particularly clear illustration of this mechanism. Although these two parameter sets belong to the consensus regime and yield approximately similar numbers of INs, they preferentially select agents with different degree compositions. In case F, the influential neighbors are more strongly weighted toward highly connected agents, which have been assigned the more positive initial belief. The resulting consensus therefore becomes increasingly positive as the number of radical hubs increases. In case G, in contrast, the threshold conditions favor a larger fraction of lower-degree INs. Because these agents are predominantly associated with the negative side of the initial belief distribution, the final consensus can become negatively biased despite the positive initial bias assigned to the hubs. Thus, increasing the number of positively biased hubs does not necessarily determine the direction of the final consensus. Rather, the direction of propagated influence depends on which agents are selected as influential neighbors by the joint trust--distrust conditions. This result highlights an important distinction between an initial bias in the belief distribution and the effective bias of the relationships through which that belief is propagated.

The comfort-index analysis provides a complementary interpretation of these numerical observations. The fraction $F$ quantifies the probability that a relationship satisfies the two influence conditions, while the comfort index $\kappa_i$ translates this probability into the expected number of INs for an agent with a given out-degree. At the population level, $\kappa_A/N=F\langle k^{out}\rangle$ therefore connects the statistical properties of trust and distrust, the threshold conditions, and network connectivity to the expected availability of influential relationships. The close correspondence between the comfort-index landscapes and the consensus--fragmentation patterns obtained from direct simulations is consistent with the interpretation that influential-neighbor availability plays an important role in consensus dynamics. However, because both the analytical landscapes and the simulations are based on the same generative trust--distrust statistics, this correspondence does not constitute an independent validation of the comfort index. Validation using empirical trust--distrust distributions would further strengthen this interpretation.

The analytical treatment also clarifies the geometric origin of this dependence. In the absence of noise, the admissible trust--distrust relationships collapse onto a deterministic line in the $T$--$U$ plane, and the transition boundary follows from the intersection of this line with the interaction territory defined by the two influence conditions. Finite noise broadens the distribution around this deterministic relation, allowing a nonzero fraction of relationships to satisfy the influence conditions over a wider range of threshold values. The resulting gradual variation of the comfort index provides a geometric explanation for the broadening and shifting of the consensus--fragmentation boundaries observed in the simulations. The comfort index therefore serves as a bridge between the microscopic statistics of trust and distrust and the macroscopic collective dynamics.

\subsection*{Limitations and future directions}

Several limitations of the present model should be noted. First, the acquaintance network and the trust--distrust relationships are assumed to
remain fixed during the belief dynamics. In real social systems, beliefs and relationships may coevolve, with changes in beliefs potentially modifying subsequent trust and distrust. Second, the analytical expression for $F$ is obtained by neglecting the truncation associated with the rejection of generated distrust values outside the interval $[0,1]$.
Consequently, the comfort-index expression provides an analytical approximation to the simulated relationship statistics rather than an exact
description of the truncated distribution. Third, the comfort index characterizes the expected availability of INs but does
not by itself determine the global equilibrium state. Consensus depends not
only on the expected number of influential relationships but also on their collective connectivity and arrangement within the network. Finally, although the sensitivity analysis in Appendix~\ref{App0} examines variations in both network size and mean out-degree, a more systematic analysis in which $N$ and $\langle k^{out}\rangle$ are varied independently would be required to disentangle finite-size effects from the effects of network connectivity.

These limitations point to several directions for future work. Allowing trust and distrust to coevolve with beliefs could provide a more realistic description of adaptive social systems and could reveal feedback mechanisms
between belief formation and relationship evolution. It would also be interesting to introduce degree-dependent or state-dependent influence
rules, which could amplify the distinction between the availability of influential neighbors and their degree composition. More systematic
analyses of network size, mean degree, and degree heterogeneity could further clarify the conditions under which network topology affects the equilibrium phase boundary. Finally, extending the present framework to time-dependent or empirically inferred trust--distrust correlations could provide a route toward connecting the mechanistic description developed
here with data-driven models of belief propagation.

\section{\label{secConclusion}Conclusion}

In conclusion, our results establish a direct mechanistic link between the joint organization of trust and distrust relationships and collective belief dynamics. The correlation between trust and distrust determines the set of admissible relationships and, together with the influence thresholds, controls the availability of influential neighbors through which beliefs can propagate. This availability provides a microscopic basis for the emergence of consensus or fragmentation, while the degree composition and initial beliefs of the INs further determine the direction and magnitude of the final consensus. Thus, the collective state cannot be understood from the marginal levels of trust or distrust, network connectivity, or initial belief distribution alone; rather, it emerges from the interplay between the statistical structure of relationships, the availability and identity of influential connections, and the propagation of beliefs through the network. The comfort index provides a compact link between these microscopic relational properties and macroscopic collective behavior, offering a framework for connecting trust--distrust structure to the emergence of collective states in networked populations.

%% file: 070_appendix.tex
\section{\label{secAppendices}Appendices }
%===========================

\subsection{\label{App0} Phase diagrams for larger network size}

\begin{figure*}[t!]
	\includegraphics[scale=0.6]{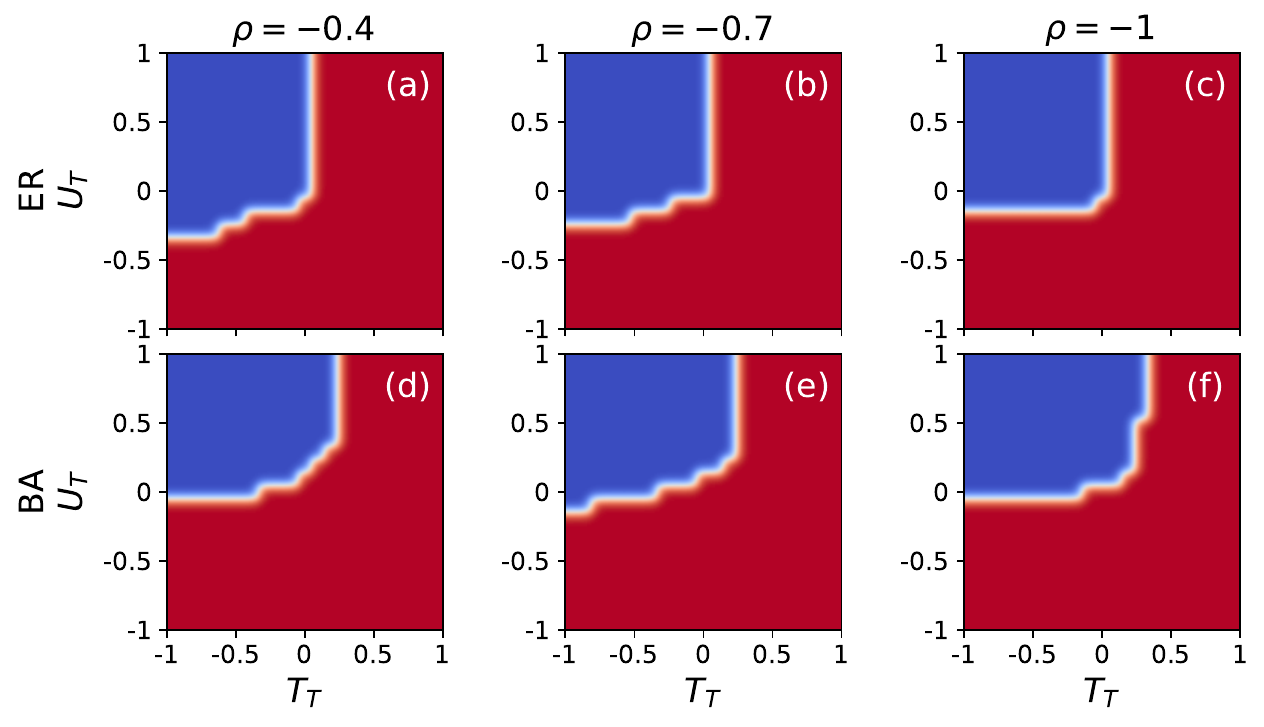}
	\centering
	\caption{\textbf{Equilibrium phase diagrams for a larger network population.}
		Results are shown for $N=5000$, mean out-degree $\langle k^{out}\rangle=50$, and $\sigma_\epsilon=0.15$.
		Panels (a)--(c) correspond to Erd\H{o}s--R\'enyi (ER) networks, while panels (d)--(f) correspond to Barab\'asi--Albert (BA) networks.
		From left to right, the magnitude of the negative correlation between the trust degree $\tau_{ji}$ and distrust degree $\delta_{ji}$ increases, with $\rho=-0.4$, $-0.7$, and $-1$.}
	\label{fig:N5000}
\end{figure*}

To examine the robustness of the equilibrium phase diagrams to network size, we repeated the simulations for larger networks with $N=5000$ and mean out-degree $\langle k^{out}\rangle=50$. The resulting phase diagrams for the Erd\H{o}s--R\'enyi and Barab\'asi--Albert networks are shown in Fig.\ref{fig:N5000}. Compared with the corresponding cases in Fig.~\ref{phase15}, which use $N=1000$ and $\langle k^{out}\rangle=100$, the overall structure of the consensus--fragmentation boundary and its dependence on the trust--distrust correlation remain qualitatively similar for both network structures. Some quantitative shifts in the phase boundaries are nevertheless observed, reflecting the combined change in network size and mean degree. These results indicate that the main qualitative features reported in the main text persist for a substantially larger network population.

%%%%%%%%%%%%%%%%%%%%%%%%%%%	
%%%%%%%%%%%%%%%%%%%%%%%%%%%	

\subsection{\label{AppA} Geometric boundary for influential-neighbor availability in the absence of noise}

\begin{figure*}[t!]
	\includegraphics[scale=0.65]{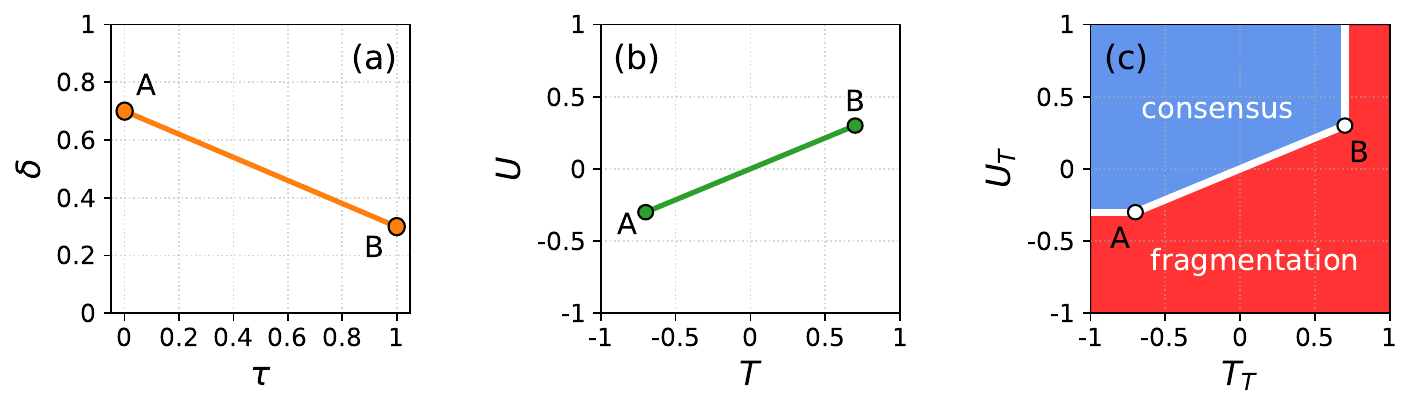}
	\centering
	\caption{\textbf{Analytical boundary for influential-neighbor availability in the absence of noise.}
		(a) Valid trust and distrust degrees for a given correlation $\rho$. All $(\tau,\delta)$ values lie on the line connecting A and B, given by Eq.~(\ref{corrEquCase1}).
		(b) Corresponding net trust and uncertainty degree in the $T-U$ plane.
		(c) Analytical phase diagram in the $(T_T,U_T)$ plane. The white curve indicates the geometric boundary for influential-neighbor availability. For the complete graph in the thermodynamic limit, this boundary is both necessary and sufficient for INs to exist, increasing the likelihood of the blue and red regions represent consensus and fragmentation, respectively. Points A and B denote the endpoints of the admissible line segment in the $T-U$ plane.}
	\label{phaseCurveFig}
\end{figure*}

In this section, we first consider the limiting case of vanishing noise, $\sigma_{\epsilon}=0$, for which the trust and distrust degrees satisfy a deterministic linear relation. From Eq.~(\ref{corrEqu}), with $\mu_{\epsilon}=\sigma_{\epsilon}=0$, we have
\begin{equation}
	\delta = \rho (\tau - 0.5) + 0.5.
	\label{corrEquCase1}
\end{equation}
For $-1\leq\rho\leq0$, this relation is a line connecting points A and B in the $\tau-\delta$ plane, as illustrated in Fig.~\ref{phaseCurveFig}(a).

Using Eqs.~(\ref{TEqu}) and (\ref{UEqu}), the corresponding net trust and uncertainty degree are
\begin{eqnarray}
	T = (1-\rho) (\tau-0.5)
	\label{Tequ} \\
	U = (1+\rho) (\tau-0.5) 
	\label{Uequ}
\end{eqnarray}
Therefore, the admissible values of $T$ and $U$ lie on the straight line
\begin{equation}
	U = \frac{1+\rho}{1-\rho}T.
	\label{UTequ}
\end{equation}
For $-1\leq\rho\leq0$, the slope of this relation is non-negative and does not exceed unity. Thus, although the trust and distrust degrees are negatively correlated in the $\tau-\delta$ plane, the corresponding net trust and uncertainty degree are positively related in the $T-U$ plane. To characterize the boundary for the existence of influential neighbors, we first obtain the coordinates of the endpoints A and B in the $T-U$ plane. Since $\tau,\delta\in[0,1]$, the endpoints in the $\tau-\delta$ plane are
\begin{eqnarray}
	(\tau_A,\delta_A)=\left(0,\frac{1-\rho}{2}\right), \\
	(\tau_B,\delta_B)=\left(1,\frac{1+\rho}{2}\right).
\end{eqnarray}
Consequently,
\begin{eqnarray}
	T_A = \tau_A-\delta_A
	= -\frac{1-\rho}{2}, \\
	U_A = \tau_A+\delta_A-1
	= -\frac{1+\rho}{2},\\
	T_B = \tau_B-\delta_B
	= \frac{1-\rho}{2}, \\
	U_B = \tau_B+\delta_B-1
	= \frac{1+\rho}{2}.
\end{eqnarray}
%%%%%%%%%%%%%%%%%%%%%%%%%%
%%%%%%%%%%%%%%%%%%%%%%%%%%
The influence conditions require $T>T_T$ and $U<U_T$. Therefore, if $U_T\leq U_A$, the interaction territory does not intersect the admissible line segment and no IN can exist. Similarly, if $T_T\geq T_B$, the interaction territory does not intersect the line segment regardless of $U_T$. Between these two limiting cases, the boundary of the region in which influential neighbors can exist is obtained when the threshold point \((T_T,U_T)\) lies on the line connecting A and B. Hence, the corresponding geometric boundary is
\begin{equation}
	\left\{ \begin{array}{l}
		{U_T^c} = U_A = - \frac{1+\rho}{2},~~~~~~~~~~ T_T \le T_A \\
		{U_T^c} = \frac{1+\rho}{1-\rho} T_T,~~~~~~~~~~~~ T_A \le T_T \le T_B\\
		T_T^c = T_B = \frac{1-\rho}{2},~~~~~~~~~~~~ T_T \ge T_B 
	\end{array} \right.
	\label{curveEpsilon0}
\end{equation}
This piecewise relation gives the analytical boundary separating the parameter region in which the interaction territory intersects the admissible set of trust and uncertainty values from the region in which no influential neighbor can exist. For homogeneous networks, this geometric boundary provides a necessary condition for the existence of influential neighbors in the absence of noise. The existence of an intersection between the admissible set and the interaction territory indicates that influential neighbors can exist, but does not by itself guarantee global consensus, which also depends on network connectivity and the initial belief configuration.

\subsection{\label{AppB} Calculating the comfort index}
	
We next derive the comfort index by obtaining the joint density function (JDF) of net trust and uncertainty degree in the $T-U$ plane. For an agent $i$ with out-degree $k_i^{out}$, the comfort index is defined as the expected number of its outgoing neighbors that satisfy the two influence conditions. We start from the generative model for the trust and distrust degrees. Defining
\begin{equation}
	x=\tau-0.5,
\end{equation}
the model in Eq.~(\ref{corrEqu}) can be written as
\begin{equation}
	x\sim \mathcal{U}(-0.5,0.5),
	\qquad
	\epsilon\sim \mathcal{N}(0,\sigma_\epsilon^2),
\end{equation}
and
\begin{equation}
	\delta-0.5=\rho x+\epsilon.
	\label{deltaCentered}
\end{equation}
The variables $x$ and $\epsilon$ are independent, and their joint density is therefore
\begin{equation}
	f_{x,\epsilon}(x,\epsilon)
		= \frac{1}{\sqrt{2\pi}\sigma_\epsilon}
		\exp\left(-\frac{\epsilon^2}{2\sigma_\epsilon^2}\right),
		\qquad
		-\frac12\leq x\leq\frac12.
		\label{jointxe}
\end{equation}	
Using the definitions of net trust and uncertainty degree,
\begin{equation}
		T=\tau-\delta,
		\qquad
		U=\tau+\delta-1,
\end{equation}
and Eq.~(\ref{deltaCentered}), we obtain
\begin{eqnarray}
	T = (1-\rho)x-\epsilon,
	\label{Tfromxeps}\\
	U = (1+\rho)x+\epsilon.
	\label{Ufromxeps}
\end{eqnarray}
The inverse transformation is
\begin{equation}
		x = \frac{T+U}{2},
		%\label{xFromTU}
		\qquad
		\epsilon =
		\frac{(1-\rho)U-(1+\rho)T}{2}, 
		%\label{epsFromTU}
\end{equation}
and the Jacobian of this transformation is
\begin{equation}
		\left|
		\frac{\partial(x,\epsilon)}
		{\partial(T,U)}
		\right|
		= \frac12.
		\label{JacobianTU}
\end{equation}
Consequently, the joint density in the $T-U$ plane is
\begin{equation}
		f_{T,U}(T,U)
		= \frac{1}{2\sqrt{2\pi}\sigma_\epsilon}
		\exp\left[
		-\frac{
			\left((1+\rho)T-(1-\rho)U\right)^2
		}{
			8\sigma_\epsilon^2
		}
		\right],
		\label{JDFTU}
\end{equation}
for values satisfying
\begin{equation}
		|T+U|\leq1,
		\label{TUstrip}
\end{equation}
which follows from the constraint $\tau\in[0,1]$. Equation~(\ref{JDFTU}) describes the distribution generated by Eq.~(\ref{corrEqu}) without imposing the additional restriction $\delta\in[0,1]$. In the simulations, samples for which $\delta$ falls outside this interval are rejected and resampled. We neglect this rejection step in the analytical calculation to obtain a tractable approximation. 

For given thresholds $T_T$ and $U_T$, an outgoing neighbor is influential when $T>T_T$ and $U<U_T$. Thus, the expected fraction of outgoing neighbors that satisfy the influence conditions is obtained by integrating the JDF over the corresponding interaction territory:
\begin{equation}
	F(T_T,U_T,\rho,\sigma_\epsilon)
	= \int_{T_T}^{1}
	\int_{-1}^{U_T}
	f_{T,U}(T,U)
	\mathbb{I}(|T+U|\leq1)
	dUdT,
	\label{Fequsec}
\end{equation}
where $\mathbb{I}(\cdot)$ is the indicator function of the admissible region implied by $\tau\in[0,1]$. The analytical expression for $F$ should be interpreted as an approximation because the rejection procedure used in the simulations is not included in the derivation above. The approximation is expected to be most accurate for weak-to-moderate negative correlations and small noise amplitudes, for which the rejection of samples with $\delta\notin[0,1]$ has a relatively limited effect on the resulting distribution of $(T,U)$. For strongly negative correlations combined with larger values of $\sigma_\epsilon$, the rejection procedure can substantially modify the empirical distribution relative to the untruncated Gaussian construction, and the analytical estimate of $F$ may therefore deviate from the fraction observed in simulations.
	
Since each of the $k_i^{out}$ outgoing neighbors is sampled independently from the same trust--distrust distribution, the comfort index, defined as the expected number of influential neighbors, is
\begin{equation}
		\kappa_i = k_i^{out}
		F(T_T,U_T,\rho,\sigma_\epsilon). \nonumber
\end{equation}
Thus, $\kappa_i$ is an expected number of influential neighbors rather than a guaranteed count. In particular, $\kappa_i\ll1$ indicates that having no influential neighbor is likely, whereas $\kappa_i\approx1$ corresponds to an expected number of influential neighbors of order unity.
	
Substituting Eq.~(\ref{JDFTU}) into Eq.~(\ref{Fequ}), we obtain
\begin{equation}
	\begin{aligned}
		&F(T_T,U_T,\rho,\sigma_\epsilon)
		= \frac{1}{2\sqrt{2\pi}\sigma_\epsilon}
		\int_{T_T}^{1}\int_{-1}^{U_T}
		\\
		&\quad \times
		\exp\left[
		-\frac{
			\left((1+\rho)T-(1-\rho)U\right)^2
		}{
			8\sigma_\epsilon^2
		}
		\right]
		\mathbb{I}(|T+U|\leq1)
		\,dU\,dT.
	\end{aligned}
	\label{FequFinal}
\end{equation}
For the noise-free case, $\sigma_\epsilon=0$, the comfort index can be obtained explicitly from the deterministic relation between $T$ and $U$. From Eqs.~(\ref{Tfromxeps}) and (\ref{Ufromxeps}), we have
$$
T=(1-\rho)x,\qquad U=(1+\rho)x,
$$
where $x=\tau-0.5$ is uniformly distributed over $[-0.5,0.5]$. The two influence conditions, $T>T_T$ and $U<U_T$, therefore impose
$$
x>\frac{T_T}{1-\rho},
\qquad
x<\frac{U_T}{1+\rho}.
$$
Consequently, for $\rho>-1$, the fraction of edges satisfying both influence conditions is given by the length of the corresponding interval within the allowed range of $x$:
$$
F_0=
\max\left[
0,\,
\min\left(\frac{1}{2},\frac{U_T}{1+\rho}\right)
-
\max\left(-\frac{1}{2},\frac{T_T}{1-\rho}\right)
\right].
$$
Here, the interval has unit total length because $x$ is uniformly distributed over $[-0.5,0.5]$, so its length directly gives the fraction of edges satisfying the influence conditions. This expression is equivalent to the geometric construction described in Appendix~\ref{AppA}, where $F_0$ corresponds to the fraction of the deterministic $T$--$U$ line lying within the interaction territory.